\documentclass[pre,final,twocolumn,showpacs]{revtex4}

\usepackage{graphicx}
\usepackage{dcolumn}
\usepackage{bm}

\begin{document}

\title{First evidence of anisotropic quenched disorder effects on a smectic liquid crystal confined in porous silicon.}

\author{R\'egis Gu\'egan}
\author{Denis Morineau}
\email{denis.morineau@univ-rennes1.fr}

\author{Claude Loverdo}
\author{Wilfried B\'eziel}
\affiliation{Groupe Mati\`ere Condens\'ee et Mat\'eriaux, UMR-CNRS 6626, Universit\'e de Rennes 1, 35042 Rennes, France}

\author{Mohammed Guendouz}
\affiliation{Laboratoire d'Optronique, FOTON, UMR-CNRS 6082, 22302 Lannion, France} 

\date{\today}

\begin{abstract}
We present a neutron scattering analysis of the structure of the smectic liquid crystal octylcyanobiphenyl (8CB) confined in one-dimensional nanopores of porous silicon films (PS). The smectic transition is completely suppressed, leading to the extension of a short-range ordered smectic phase aligned along the pore axis. It evolves reversibly over an extended temperature range, down to 50 K below the \textit{N-SmA} transition in pure 8CB. This behavior strongly differs from previous observations of smectics in different one-dimensional porous materials.
A coherent picture of this striking behavior requires that quenched disorder effects are invoked. The strongly disordered nature of the inner surface of PS acts as random fields coupling to the smectic order. The one-dimensionality of PS nano-channels offers new perspectives on quenched disorder effects, which observation has been restricted to homogeneous random porous materials so far.

\end{abstract}
\pacs{61.12.Ex, 61.30.Eb, 64.70.Md,61.30.Pq}
\maketitle

\section{introduction}
The smectic phase transition of liquid crystals (LC's) confined in porous geometry have recently retained much attention \cite{Bellini-Science-01}. It offers remarkable opportunities to address fundamental questions arising from statistical mechanics that challenge current theories. In this field, the effect of quenched disorder on a continuous symmetry breaking transition is an especially interesting topic. Smectic materials are elastically soft and may directly couple to the surface of the porous matrix. Indeed, smectic confined in random porous aerogels are definitively recognized as exceptional model systems \cite{Bellini-PRL-93}. The companion experimental test-systems currently accessible are (LC)-aerosil dispersions \cite{Leheny-PRE-03}. They have been shown to provide access to the regime of weak disorder with a direct control of its strength.

Major outcomes have been provided by scaling analysis of X-Ray scattering experiments and heat capacity measurements \cite{Leheny-PRE-03,Clegg-PRE-03,Bellini-Science-01}. General common features have been observed for these two model systems, although some still-questioning differences remain. The most striking observation is the absence of a true nematic-to-smectic transition, which is replaced by the gradual occurrence of a short-range ordered phase. This confirms the general theoretical prediction that a transition breaking a continuous symmetry is unstable toward arbitrarily small quenched random fields. A detailed theoretical analysis of the effects of disorder on the stability of the smectic phase at low temperature and the elasticity of the phase has been proposed recently \cite{Radzihovsky-PRE-99}. It leads to various predictions that have been mostly verified experimentally in the case of aerogels. The situation is slightly different for aerosils, where discrepancies have been reported. This difference is not obviously related to the strength of disorder. Indeed, the theory is expected to be valid for weak disorder, which better describes the case encountered for aerosils.

Both aerogels and aerosils structures are disordered at long distances and spatially isotropic. It leads to an almost spatially homogeneous random pinning, which couples directly to the smectic order and nematic director. Beyond these isotropic geometries, there is a growing interest in the creation of anisotropic random media as they are expected to reveal new perspectives of the subject \cite{Leheny-JPCM-04,Schoenhals-CPL-99}. A better understanding of the disorder effects induced by porous geometries could be achieved by the disentangling of random fields coupling to the nematic and the smectic order, respectively. In addition, the low temperature topologically ordered glass phase predicted for smectics confined in anisotropic random media is a totally open problem from the experimental point of view \cite{Radzihovsky-PRL-99}. Anisotropic homogeneous random media have been recently obtained from strained colloidal silica gels (aerosils) \cite{Leheny-JPCM-04}. In the present contribution, we are able to introduce anisotropic random fields in 1-Dimensional conditions of confinement using aligned nanochannels formed in porous silicon films (PS). We prove that the highly corrugated inner surface of PS directly disrupts the smectic ordering of 8CB, while preserving a long-range orientational order.

\begin{figure*}[tbp]
\includegraphics[width=0.85\hsize]{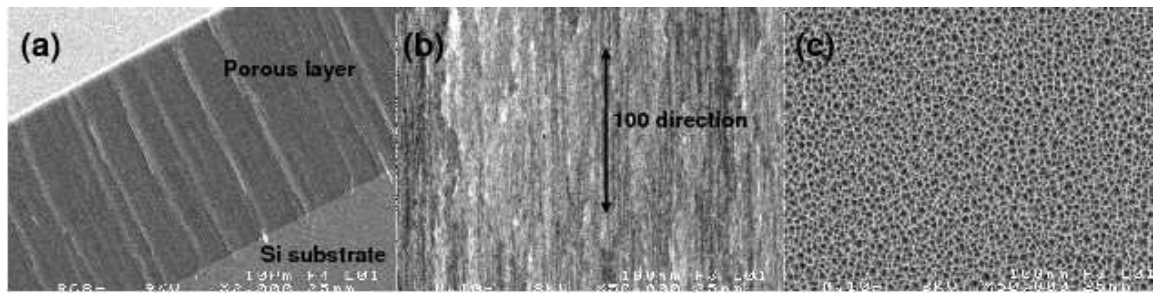}
\caption{\label{FIG-Meb} Scanning electron micrographs of the porous silicon film. (a) side view at low magnification showing the 30 $\mu$m thick porous layer attached to the silicon substrate. (b) side view at higher magnification showing the aligned mesoporous channels. An irregular inner pore morphology is typically obtained under the electrochemical conditions used in this study, as detailed in Ref. \cite{Lehmann-MS-00}. (c) top view at higher magnification.  }
\end{figure*}

\section{Experimental}
\subsection{Sample}
PS layers are obtained by electrochemical anodization of (100)-oriented silicon substrates in an HF electrolyte solution. All the properties of PS layers, such as porosity, thickness, pore diameter, microstructure are strongly dependent on the anodization conditions and have been extensively studied \cite{Lehmann-MS-00}. Anodization of heavily doped (100)-oriented silicon leads to highly anisotropic porous layers consisting of macroscopically long channels (diameter $\sim$ 100 \AA) running perpendicular to the surface wafer (also denoted as the columnar form of PS) as shown in Fig.~\ref{FIG-Meb}. Different morphologies are obtained for lightly doped (p-type or n-type) silicon, leading to various degrees of branching and eventually to isotropic microstructures \cite{Lehmann-MS-00}. The porous samples we used were obtained from heavily p+ doped (100)-silicon substrate in electrochemical conditions, which provide the columnar form of PS, as detailed elsewhere \cite{Guendouz-PSS-03}. The porous geometry can be described as a parallel arrangement of strongly anisotropic not-connected channels (diameter 100 \AA, length 30 $\mu$m). 

This porous volume is considered as highly regular since its orientational order extends to macroscopic lengthscales. This is strictly valid for each individual channel, which presents a straight main axis of several tens of micrometers in length. The aspect ratio of each individual channel typically exceeds 3000:1 and induces a lower dimensionality (quasi-1D) to the system. In addition, all the channels are aligned perpendicularly to the silicon surface, which is about 1cm$^2$. This second feature reduces powder average effects and allows measuring the anisotropic components of each unidimensional nanoconfined system from experiments that probe the entire porosity. On the other hand, the inner surface of the 1-D pores is characterized by a significant disorder at microscopic lengthscales  i.e. $\sim$1 nm. This third feature of the columnar form of PS is reminiscent of other types of PS morphologies obtained when decreasing the doping density or changing the anodization conditions. This exemplifies a tendency of porous silicon films to present eventually branching, microporosity or isotropic porosity. A systematic description of the various porous morphologies of PS has been reported recently by Lehmann et al. \cite{Lehmann-MS-00}.

The combination of macroscopic order versus nanoscopic disorder in PS is unique among porous geometries, and is essential to the aim of this study. The inner pore surface proceeds as a random disruption of the translational order related to the smectic phase. However, at variance with homogeneous media such as aerogels and aerosils, the long-range ordering of these local random fields leads to a long-range orientational order of the confined phase. Large quenched disorder effects induced by the morphology of PS have been recently invoked by Wallacher \textit{et al.} for the capillary condensation \cite{Knorr-PRL-04}. It should be also noticed that PS presents a native oxide layer that covers the entire inner surface of the pore. It has been intentionally preserved for this study, so that the interfacial interaction energy between 8CB and the PS compares to the one obtained with other porous silica matrices. To summarize, the main differences between PS and previously studied porous materials such as aerogels essentially arise from a significantly smaller pore size, a stronger induced disorder and an anisotropic morphology.

\subsection{Neutron scattering}
Prior to neutron scattering experiments, outgased 8CB (spectroscopic grade) was confined into PS wafers by impregnation from the liquid phase for a duration of two hours. During this procedure, the temperature of the sample was kept at 60 degrees Celsius in a vacuum chamber maintained under 8CB vapor pressure. The saturated sample was wiped clean of excess liquid crystal. In order to increase the signal of diffraction, eight PS wafers were placed in a holder that imposes a parallel alignment of the porous channels and placed in a laboratory made aluminum cylindrical cell. The neutron scattering experiments were performed on the double-axis spectrometer G6.1 of the Laboratoire L\'eon Brillouin neutron source facility (CEA-CNRS, Saclay) using a monochromatic incident wavelength of 4.7 \AA. A thermostatic fluid circulation loop was used to heat/cool the sample, which temperature was controlled with a stability better than 0.1 K. The measurement temperatures ranged from 255 K to 310 K, from $\sim$ 52 K  below to $\sim$ 3 K above the \textit{N-SmA} transition in pure 8CB.

By a simple rotation of the cylindrical cell, it is possible to change the angle between the pores axis and the incident beam. Two peculiar configurations, corresponding to grazing and normal incidences have been selected so that the transfer of momentum \textbf{q}, in the limit of small angles of diffraction, is practically parallel and perpendicular to the pores axis, respectively.

\section{Results}
\begin{figure}[btp]
\includegraphics[width=0.85\hsize]{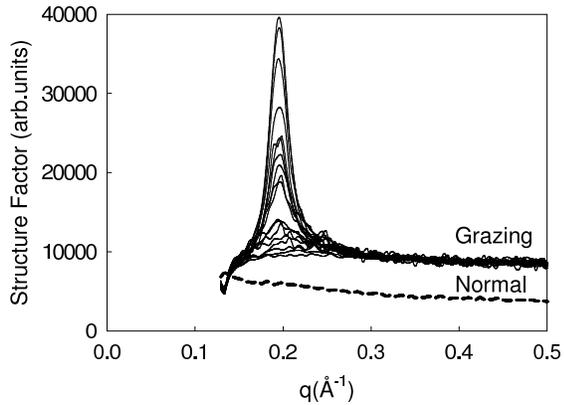}
\caption{\label{FIG-Spectra} Progressive growth of the smectic peak in the neutron structure factor of 8CB confined in porous silicon. Temperatures range from 310 K to 255 K. These spectra have been obtained in grazing incidence. No smectic peak is observed in normal incidence at 255 K (dashed line shifted down for clarity), which demonstrates the parallel axial alignment of the phase.}
\end{figure}

Below the bulk smectic transition ($T^0_{NA}=306.7 K$), a Bragg peak located around $q=0.2 $ \AA$^{-1}$ {} has been observed in grazing incidence only, as shown in Fig.~\ref{FIG-Spectra}. Conversely, the spectra obtained in normal incidence present diffuse scattering only. This demonstrates that the confined phase develops an orientational order parallel to the pore axis. Neutron experiments using a 2D multidetector and ellipsometry have confirmed the preferential alignment of the smectic phase. The width of the angular distribution of the smectic direction is $30^{\circ}$ and does not show significant variation in the temperature range covered by this study \cite{Morineau}.

The second striking feature shown in Fig.~\ref{FIG-Spectra} is the complete destruction of the \textit{N-SmA} transition. Indeed, the Bragg peak related to the smectic translational order appears progressively from about the bulk transition temperature $T^0_{NA}$ down to the lowest temperature accessible with no sign of crystallization $\sim$255 K. Even at this temperature, the peak intensity does not saturate, indicating that the smectic order continuously and reversibly evolves over more than 50 K. The extreme depression of the crystallization point of 8CB in PS offers an unique opportunity to study this short-ranged smectic phase at equilibrium over an extended temperature range. As a comparison, the study of this phase is limited to 273 K and 287 K respectively for 8CB confined in aerogels and aerosils, i.e. very close to the bulk crystallization itself \cite{Leheny-PRE-03,Bellini-Science-01}.

\begin{figure}[tbp]
\includegraphics[width=0.85\hsize]{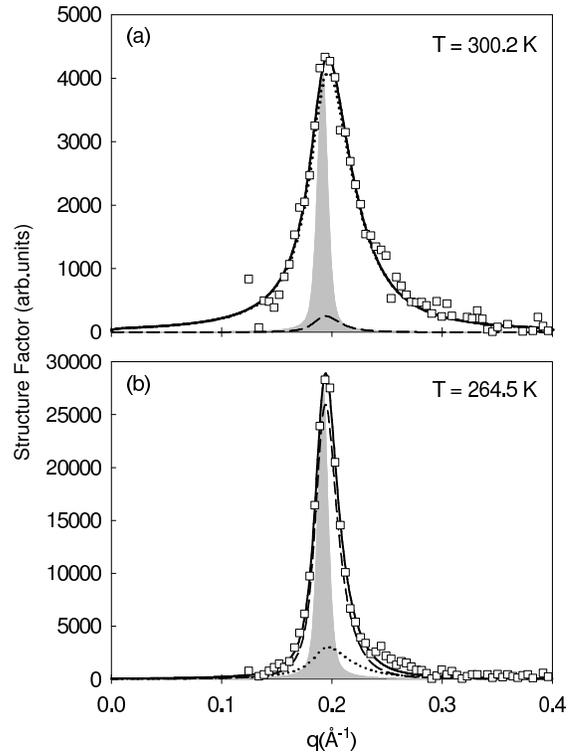}
\caption{\label{FIG-spectra-fit} Neutron structure factor of 8CB confined in porous silicon in the region of the smectic Bragg peak after subtraction of the background measured at 315 K (open squares). The spectrometer resolution (shaded area) highlights the  broadening of the peak. The full line results from the analysis with a two components model including a thermal term (dotted line) and a disorder term (dashed line). The intensity arises essentially from thermal fluctuations at 300.2 K, whereas it is dominated by the random-field contribution at 264.5 K. Note the difference in vertical scales between the two graphs.}
\end{figure}

Furthermore, the measured smectic peak remains significantly broader than the experimental resolution, even at the lowest temperatures (cf. Fig.~\ref{FIG-spectra-fit}). For pure 8CB, the quasi-long-range ordered smectic phase is characterized by a resolution-limited Bragg peak. This demonstrates that confinement in PS leads to a finite extension of the smectic correlations.

\section{Analysis and discussion}

We have analyzed the shape of the Bragg peak using the formalism derived from random fields theories and successfully applied to LC confined in random homogeneous porous materials. After subtraction of a diffuse background, which has been obtained experimentally from the spectrum of the isotropic phase at 315 K, the scattered intensity in the region of the Bragg peak arises from two contributions as written in Eq.~(\ref{eq1}). 

\begin{eqnarray}
S(\mathbf{q})=\frac{\sigma_1}{1+\left(q_{\|}-q_0\right)^2\xi_{\|}^2+q_{\bot}^2\xi_{\bot}^2+cq_{\bot}^4\xi_{\bot}^4} 
\nonumber \\
+\frac{a_2\left(\xi_{\|}\xi_{\bot}^2\right)}{\left[1+\left(q_{\|}-q_0\right)^2\xi_{\|}^2+q_{\bot}^2\xi_{\bot}^2+cq_{\bot}^4\xi_{\bot}^4\right]^2}
\label{eq1}
\end{eqnarray}

The first term is due to thermal fluctuations, whereas the second reveals the effects of quenched disorder. $\xi_{\|}$ and $\xi_{\bot}$ are the smectic correlation lengths in the direction parallel and perpendicular to the smectic wave vector $\textbf{q}_0$, respectively. This expression has been integrated numerically over the angular distribution of the smectic orientation ($\theta_m=30^{\circ}$) and convoluted to the experimental resolution $R(q)$ obtained by calibration with a standard crystal (cf. Eq.~(\ref{eq2})).

\begin{eqnarray}
I(q)=\frac{1}{1-\cos(\theta_m)}\int dq'\int^{\theta_m}_{0}R(q-q')S(\mathbf{q})\sin(\theta)d\theta
\label{eq2}
\end{eqnarray}

\begin{figure}[tbp]
\includegraphics[width=0.85\hsize]{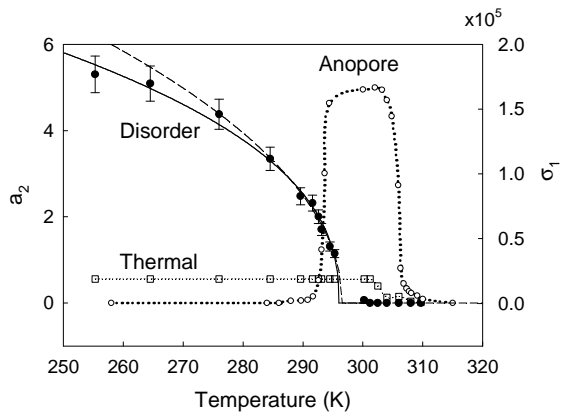}
\caption{\label{FIG-Intensity} Temperature variation of the integrated area $a_{2}$ of the static fluctuation term (filled circles) and  the amplitude of the thermal fluctuations $\sigma_1$ (open squares) of 8CB confined in porous silicon. The solid and dashed lines are two power law fits of $a_{2}$ (see text). These features differ from 8CB confined in porous alumina (Anopore), whose Bragg peak intensity shows well-defined nematic-to-smectic and crystallization transitions (open circles with dotted line), which are only slightly depressed from the bulk.}
\end{figure}

The use of this formula allowed a satisfactory fit of the data in the whole smectic temperature range. A first fitting procedure has been conducted with every parameter free and independent. It appeared that the introduction of the parameter \textit{c}, which corresponds to an empirical $q^4$ correction to the Lorentzian function, did not improve significantly the quality of the adjustment. It is therefore impossible to extract a reliable value of \textit{c} within experimental uncertainties. In previous studies, some authors have firstly chosen to fix its value to zero, like for 8CB in aerogels by Bellini \textit{et al.} \cite{Bellini-Science-01} or have retained the same variation with $\xi_{\|}$ as in pure 8CB (cf. aerosils by Leheny et al. \cite{Leheny-PRE-03}). For consistency with this last analysis, the results reported in this paper have been obtained by application of the same procedure. The values of \textit{c} in the bulk have been obtained from \cite{Ocko-ZPB-86} and were of the order of 0.2 in the range of $\xi_{\|}$ covered by our analysis. We have also observed a coupling between $\xi_{\|}$ and $\xi_{\bot}$, the ratio of these two correlation lengths slightly varying from 3 to 4. These experimental data are also compatible with the assumption that $\xi_{\|}$ and $\xi_{\bot}$ follow the same relative variation as in the bulk \cite{Ocko-ZPB-86}. In order to limit the number of free parameters, we assumed the relation to be valid for 8CB in PS as previously done for aerosils \cite{Leheny-PRE-03}. Hence, the number of free parameters is limited to four : $a_2$, $\sigma_{1}$, $\xi_{\|}$ and $q_0$.

At high temperature, from 310 K to 300 K, the scattering intensity is well described by thermal fluctuations alone and the static fluctuation term $a_2$ is essentially zero. This can be attributed to pre-transitional thermal fluctuations, which in bulk lead to a diverging susceptibility at $T^0_{NA}$. For 8CB confined in porous silicon, the divergence of $\sigma_{1}$ is suppressed and the amplitude of thermal fluctuations saturates as shown in Fig.~\ref{FIG-Intensity}. Below an avoided phase transition point located around 300 K, $\sim$7 K below $T^0_{NA}$, the contribution from static fluctuations rises sharply and represents the essential of the scattering intensity. 

Leheny \textit{et al.} \cite{Leheny-PRE-03} have recently shown that the integrated area of the disorder term, which is close to $a_2$, varies with temperature like an order parameter squared in the vicinity of the pseudotransition. Obviously, the temperature variation of $a_2$ for 8CB confined in PS follows the same trend. Reasonable fits of the data can be obtained with the functional form \begin{eqnarray}
a_2\propto\left(T^*-T\right)^{2\beta},
\label{eq3}
\end{eqnarray} 
as displayed in Fig. ~\ref{FIG-Intensity}. The use of PS has allowed us to extend the domain of stability of the smectic phase to much lower temperature at the cost of a reduced number of spectra in the region of the pseudotransition. Hence, obtaining reliable values of critical exponent is probably illusive but give useful indications.    
A fit over the entire temperature range gives $T^* = 296.0$ K and a pseudo-critical exponent $2\beta$ of 0.41. Restricting the fit to a temperature range of 5 K below the transition gives $T^* = 296.6$ K and $2\beta=0.49$. These two limiting cases appear as solid and dashed lines in Fig.~\ref{FIG-Intensity}. The exponent is comparable with $2\beta=0.38$ obtained for 8CB confined in aerogels \cite{Bellini-Science-01} and $2\beta$ ranging from 0.45 to 0.65 for aerosils \cite{Leheny-PRE-03}. Although the present porous system presents significant differences from these two latter cases in terms of pore size and morphology, it stresses again the similar underlying nature of the observed phenomenon.

\begin{figure}[tbp]
\includegraphics[width=0.85\hsize]{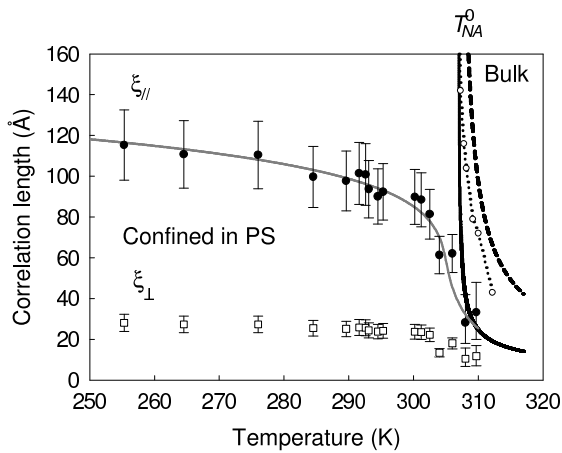}
\caption{\label{FIG-Ksi} Temperature variation of the smectic correlation lengths $\xi_{\|}$ (filled circles) and $\xi_{\bot}$ (open squares) for 8CB confined in porous silicon. The gray solid line is a guide for the eyes. The same quantities for bulk 8CB are plotted in dashed and solid lines respectively \cite{Ocko-ZPB-86}. The smectic correlation lengths $\xi_{\|}$ for 8CB confined in porous alumina (anopore) is also displayed for comparison (open circles with dotted line).}
\end{figure}

The smectic correlation length extracted by the fitting procedure is finite at every temperature (cf. Fig.~\ref{FIG-Ksi}). This was announced by the observation of a broadened Bragg peak in Fig. ~\ref{FIG-Spectra}. Above $T^0_{NA}$, the smectic correlation length $\xi_{\|}$ of confined 8CB is about $30$ \AA, which is practically the lowest meaningful value given by the molecular size. In bulk 8CB, similar values are observed in smectic thermal fluctuations a few K below the clearing point. In contrast with the bulk, $\xi_{\|}$ of 8CB confined in PS does not show any divergent behavior on approaching $T^0_{NA}$ from above. It starts to increase moderately below $T^0_{NA}$ and reaches a saturating value of $120$ \AA{} at 255 K. The small variation of $\xi_{\|}$ where disorder dominates in comparison to pretransitional thermal  fluctuations does not allow extracting a reasonable scaling law. At variance with 8CB in aerosils, the smectic correlation length of 8CB in PS does not saturate just below $T^0_{NA}$ to its low-temperature limit but displays a progressive increase along with the increase of the peak intensity. This behavior qualitatively compares the one reported for 8CB in aerogels, which displays a progressive variation with temperature of both $\xi_{\|}$ and $|\psi|^2$ although the stronger disorder induced by PS considerably limits their amplitude.

\begin{figure}[tbp]
\includegraphics[width=0.85\hsize]{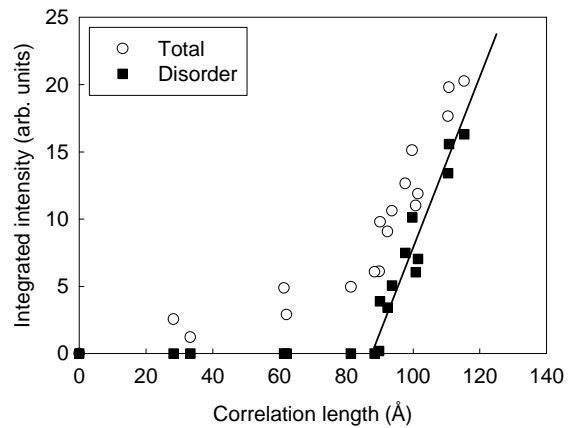}
\caption{\label{FIG-integrated-int} Integrated intensity of the smectic Bragg peak of 8CB in porous silicon as a function of the correlation length. Open circles correspond to the complete integration of the experimental structure factor, whereas solid squares correspond to the integration of the disorder component deduced from the fitting procedure.}

\end{figure}

Fig.~\ref{FIG-integrated-int} shows the variation with $\xi_{\|}$ of both the total integrated intensity of the structure factor in the region of the Bragg peak and the disorder component deduced from the previous fitting procedure. The intensity obtained by integration of the experimental Bragg peak presents two dependences with the correlation length for values laying above and below 90 \AA. This value of $\xi_{\|}$ corresponds to a temperature, which is close to $T^*$. This reflects two distinct regimes related to a different variation of the Bragg peak shape and intensity. The growth of smectic order is essentially related to thermal fluctuations above $T^*$, $\xi_{\|}$ increasing from 30 \AA{} to 90 \AA, while $a_2$ is essentially null. Below $T^*$, the saturation of the thermal fluctuations is concomitant with the onset of the static disorder term.

The linear increase of $|\psi|^2$ with $\xi_{\|}$ at low temperature resembles the behavior of 8CB in aerogels, although it is limited to much smaller values. This dependence had been reported by Bellini \textit{et al.} \cite{Bellini-Science-01} as an argument in favor of the theory of Radzihovsky and Toner \cite{Radzihovsky-PRE-99}. Indeed, the random field theory predicts a relation between $\xi_{\|}$ and the smectic elastic constant $B(T)$ as $\xi_{\|}\sim B(T)^{1/\kappa}$. Sufficiently far below $T^*$, the mean field approximation should apply so that $B(T)\sim|\psi|^2$, where $|\psi|^2$ is the smectic order parameter squared experimentally obtained by integrating the scattering intensity \cite{Radzihovsky-PRE-99}. This gives $\xi_{\|}\sim|\psi|^{2/\kappa}$, which in the absence of anomalous elasticity leads to $\xi_{\|}\sim|\psi|^2$. The validity of this scaling law has been verified for 8CB in aerogels with indications of anomalous elasticity at low temperature \cite{Bellini-Science-01}. At variance with this, 8CB in aerosils dispersions do not show any significant temperature dependence of  $\xi_{\|}$ below $T^*$ \cite{Leheny-PRE-03}. 

The interpretation of the linear dependence of the disorder susceptibility with $\xi_{\|}$ in the case of PS is not clear at present. Indeed PS introduces a large strength of disorder, so that the variation of $\xi_{\|}$ in the temperature region where the disorder term dominates is not significantly larger than the smallest thermal fluctuations already present in the pretransitional regime. In addition, the case of PS probably falls beyond the limits of validity of this prediction, which has been expressed for weak disorder.

Using 8CB in PS allows to investigate experimentally the regime of strong disorder. As a consequence, $\xi_{\|}$ converges to saturating values, which are much smaller than the one observed for aerosils and aerogels, typically of several hundreds or thousands Angstrom. The reported features are however clearly different from simple finite size effects. In this case, one would expect the smectic order to increase on approaching  $T^0_{NA}$ from above, thus tracking the bulk behavior. $\xi_{\|}$ would eventually saturate at a temperature above $T^0_{NA}$, where it reaches the typical pore dimension. The case of 8CB is very different since the progressive growth of smectic order essentially occurs below $T^0_{NA}$ and saturates 50 K below $T^0_{NA}$. The necessity to introduce the two components of Eq.~(\ref{eq1}) to describe properly the line shape provides an additional report of the primacy of quenched disorder effects. Additional arguments arise from a comparison with other membranes showing unidimensional nanopores. 8CB confined in alumina 1D channels of 200 nm from Ref. \cite{Finotello-PRE-94} and recent results for an alumina membrane of diameter 20 nm nicely illustrate an alternative scenario. They have shown a strong orientational order of the confined phase induced by anchoring effects \cite{Morineau}. However, the signature of a sharp transition from nematic to smectic phase is maintained although it is slightly rounded and depressed (about 2 K) and spatially limited by finite size effects as showed in dotted lines in Figs.~\ref{FIG-Intensity} and ~\ref{FIG-Ksi}. This is also true for the crystallization of 8CB in such membranes, which is basically reminiscent of the bulk behavior as shown in Fig.~\ref{FIG-Intensity}. These features nicely illustrate dominant finite size effects but do not compare with the behavior of 8CB in PS. The properties of 8CB in PS are conferred by the unusual strongly disordering nature of the pore surface of PS, whereas alumina is known to present a fairly regular wall structure.
 
A central question concerns the nature of the disorder introduced by porous silicon. Compared to random porous media such as aerogels and aerosil dispersions, the porosity of PS films is markedly different. The use of aligned 1D-channels allows having of fully ordered porous geometry at long scale, the disorder being therefore essentially attributed to the short-range disordering nature of the pore surface. This effect is expected to be strong for PS because of the particular inner shape of the pores. This is consistent with the large quenched disorder effects reported by Wallacher \textit{et al.} \cite{Knorr-PRL-04} for the capillary condensation of gases in PS. Previous studies on LC confined in 1D-channels (such as nuclepores and anopores) have not displayed comparable disruptions of the smectic transition because of a more regular pore surface. Although an absolute evaluation of the strength of random field disorder $\Delta$ is difficult for PS, a comparative estimation with the case of aerosils can be produced. In the latter case, $\Delta$ can be varied by more than one order of magnitude by changing the aerosils density. A systematic dependence of $\xi_{\|}$ and $a_2 / \sigma_{1}$ with $\Delta$ has been obtained (cf. Fig. 11 in Ref. \cite{Leheny-PRE-03}). The values measured for PS correspond to a strength of random field disorder 3 times greater than for the denser aerosil-dispersion. A larger value of $\Delta$ together with a stronger confinement is probably responsible for the exceptionally broad temperature range where the gradual growth of the smectic order occurs.

\section{Conclusion}
Random fields effect on a continuous symmetry transition has remained one of the challenging topics of modern statistical mechanics. It has motivated many efforts toward a better understanding of the smectic transition of LC confined in random porous geometries. Only very recently, the case of anisotropic random pores has been addressed by using strained aerosils. However, reports of quenched disorder effects  have been restricted to confinement in homogeneous random porous materials so far, such as aerogels and aerosils. In addition, these systems mainly fall in the weak disorder regime. Introducing PS as an original porous materials, we have been able to address the case of macroscopically long 1-D channels where random fields are introduced by a strongly disordered inner pore morphology. 

We have tracked the smectic correlation length in 8CB confined in PS by neutron scattering experiments in a temperature range encompassing the isotropic phase down to 255K i.e. 50 K below $T^0_{NA}$. The 1-D character of each individual pore induces an preferential alignment of the confined phase, which can be monitored with restrained powder average since all the channels are parallel.      

The behavior of confined LC drastically differs from pure 8CB and 8CB confined in anopores channels of similar size. The smectic transition is completely suppressed, leading to the extension of a short-range ordered smectic phase aligned along the pore axis. It evolves reversibly over an extended temperature range, which had never been achieved with previous random pores without crystallization. 

A complete description of the smectic lineshape has required the introduction of both thermal and disorder terms, the latter being dominant at low temperature. The correlation length and the integrated intensity of the disorder term increase concomitantly when decreasing the temperature below $T^0_{NA}$. Saturating values are reached only 50 K below $T^0_{NA}$ and the smectic order remains short ranged. Despite the preferential alignment of the confined phase and the reduced dimensionality of the pore, there is no clear evidence of any significant difference between the anisotropy of the smectic correlation lengths $\xi_{\|}$ and $\xi_{\bot}$ in PS and in the bulk. Higher resolution experiments are in progress to address this issue.    

The use of PS allows original investigations since it covers the regime of anisotropic pores and strong perturbation. In principle, these experimental conditions fall out of the limits of available theories expressed for weak disorder \cite{Radzihovsky-PRE-99}. It prevents a full confrontation and may inspire future investigations. Our results already imply that much of the scenario expected for weak disorder is retained in the regime obtainable with PS.

\section{Acknowledgments}
We thank B. Toudic and R. Lefort for fruitful discussions. Experiments have been performed at the Laboratoire L\'eon Brillouin neutron source facility (CEA-CNRS, Saclay). The assistance of J.-P. Ambroise and I. Mirebeau during this project is particularly appreciated.




\end{document}